\documentclass[%
 reprint,
 amsmath,amssymb,
 aps,
 showkeys
]{revtex4-2}

\usepackage{graphicx}
\usepackage{dcolumn}
\usepackage{bm}
\usepackage{hyperref}

\usepackage{slashed}
\usepackage{amsfonts}
\usepackage{mathtools}
\usepackage{makeidx}   
\usepackage{float}
\usepackage{gensymb}
\usepackage{comment}
\usepackage{upgreek}
\usepackage{placeins}
\usepackage{epstopdf}
\usepackage{xcolor}

\begin{document}

\preprint{APS/123-QED}

\title{Searching for Long-Lived Dark Scalars with MAPP-1 and MAPP-2 at the HL-LHC}






\author{Shafakat Arifeen}
\email{shafakat@ualberta.ca}
\affiliation{Department of Physics, University of Alberta, \\ Edmonton, Alberta, T6G 2E1, Canada}

\author{James Pinfold}
\affiliation{Department of Physics, University of Alberta, \\ Edmonton, Alberta, T6G 2E1, Canada}

\author{Pierre-Philippe A. Ouimet}
\email{pierre-philippe.ouimet@uregina.ca}
\affiliation{
Department of Physics, University of Regina, \\ Regina, Saskatchewan, S4S 0A2, Canada
}

\author{Michael Staelens}
\altaffiliation[Also at ]{Instituto de F\'isica Corpuscular, CSIC--Universitat de Val\`encia, Catedr\'atico Jos\'e Beltr\'an, 2, Paterna, 46980, Val\`encia, Spain.}
\affiliation{Department of Physics, University of Alberta, \\ Edmonton, Alberta, T6G 2E1, Canada}
\affiliation{
Department of Physics, University of Regina, \\ Regina, Saskatchewan, S4S 0A2, Canada
}

\date{\today}

\begin{abstract}
Feebly interacting particles (FIPs) arise generically in dark-sector extensions of the Standard Model (SM). Among the simplest and most theoretically motivated such candidates is a dark scalar that mixes with the SM Higgs boson, acting as a mediator between the dark sector and the SM. For sufficiently small mixing, such a dark scalar manifests at collider experiments as a long-lived particle with a macroscopic decay length---a signature to which conventional general-purpose detectors have limited sensitivity. The MoEDAL-MAPP experiment is a dedicated search facility at the Large Hadron Collider (LHC) targeting long-lived FIPs as avatars of physics beyond the SM. We present $95\%$ confidence-level background-free sensitivity projections for the MAPP-1 and MAPP-2 detectors at the High-Luminosity LHC to Higgs-portal dark scalars produced via rare $B$-meson decays and exotic Higgs boson decays. Our results establish the proposed MAPP-2 detector as a complementary probe of feebly-coupled dark scalars, with sensitivity to lifetimes beyond the reach of existing collider searches.
\end{abstract}

\maketitle


\section{\label{sec:Intro}Introduction}
Over the past few decades, astrophysical and cosmological observations have provided compelling evidence for the existence of dark matter (DM), which accounts for approximately $25\%$ of the energy density of the universe~\cite{Planck2018cosmo}. However, the Standard Model (SM) contains no viable DM candidate, and despite extensive experimental efforts, DM has yet to be directly detected~\cite{Misiaszek2024}. While DM's gravitational interactions are well established~\cite{Freese2009}, its nongravitational interactions with the SM, if any, remain unknown. One well-studied possibility is that DM interacts with SM particles through the weak interaction, as in weakly interacting massive particle (WIMP) models~\cite{Jungman1996, Roszkowski2018}. The absence of experimental evidence for WIMPs~\cite{Arcadi2018}, however, motivates the exploration of alternative scenarios in which DM is more feebly coupled to the SM~\cite{Lanfranchi2021, Antel2023}.

A particularly well-motivated possibility is that DM belongs to a dark sector containing additional particles and interactions~\cite{Arina2021}. Such sectors can communicate with the SM through so-called portal interactions~\cite{Arina2021}, provided a framework in which new particles can be produced at particle colliders while remaining sufficiently weakly coupled to evade conventional searches. Among the possible portal interactions, a particularly simple and well-motivated realization is the Higgs portal~\cite{Patt2006}, where a dark-sector scalar mixes with the SM Higgs and inherits small couplings to ordinary matter as a result~\cite{Clarke2014}. Although these couplings can be small, they may still be sufficient to produce the dark scalar in SM processes and admit subsequent decays to visible particles~\cite{Ferber2024}. Notably, the couplings small enough to give macroscopic decay lengths are not excluded by current experimental limits, leaving this long-lived regime open for discovery~\cite{Ferber2024}.

Extensions of the SM containing additional scalar fields can provide new mechanisms for symmetry breaking~\cite{Ivanov2017} and alter the electroweak phase transition, with implications for the evolution of the early Universe and the generation of the baryon asymmetry~\cite{Morrissey2012}. In particular, a scalar field in the dark sector can play a role in models of inflation~\cite{Shaposhnikov2006, Bezrukov2010, Bezrukov2013, Bramante2016}, while scalar-mediated interactions between the dark and visible sectors can affect the thermal history and relic abundance of DM~\cite{Burgess2001, Arcadi2020}. These connections make dark scalars relevant to several outstanding questions in particle physics and cosmology, independently of whether the scalar itself constitutes the DM. Moreover, their potentially rich phenomenology, combined with the possibility of long lifetimes, makes dark scalars compelling targets for collider searches~\cite{Ferber2024}.

Long-lived dark scalars are therefore particularly interesting targets for experiments designed to search for feebly interacting particles (FIPs). Their suppressed interactions can allow them to travel substantial distances before decaying, producing displaced-vertex signatures that may be challenging to detect with conventional collider detectors but accessible to dedicated long-lived particle (LLP) searches. Dedicated detectors at colliders such as the Large Hadron Collider (LHC) are specifically designed to capture such displaced signatures, providing sensitivity to regions of dark-scalar parameter space that may be difficult to access with conventional collider detectors.

The MoEDAL's Apparatus for Penetrating Particles (MoEDAL-MAPP) experiment is a dedicated search experiment at the LHC designed to investigate physics beyond the Standard Model (BSM) through the detection of FIPs~\cite{pinfold2023moedalmapp}. A substantial body of work has investigated the sensitivity of the MoEDAL-MAPP experiment to a broad range of FIP scenarios~\cite{Deppisch2019, Frank2020, deVries2021, Dreiner2021, StaelensPhDThesis, Popa2022, Bertuzzo2022, Beltran2023, Dreiner2023, Kalliokoski2024, Staelens2024, Alemany2025, Kalliokoski2025, Deppisch2026, deMontigny2026, Arifeen2026, Wang2026}. Long-lived dark scalars have also been considered in the context of both MAPP-1 and MAPP-2~\cite{StaelensPhDThesis, Staelens2024, Alemany2025}, with existing sensitivity projections based exclusively on dark-scalar production in $B$-meson decays. A comprehensive study of the dark-scalar parameter space at MoEDAL-MAPP, including additional production channels relevant to the High-Luminosity LHC (HL-LHC), has yet to be published.

In this study, we present the projected sensitivities of MAPP-1 and MAPP-2 to long-lived dark scalars at the HL-LHC. Section~\ref{sec:MAPP} describes the MoEDAL-MAPP experimental program. Section~\ref{sec:HiggsPortal} introduces the Higgs-portal dark-sector model and discusses the relevant dark-scalar production and decay phenomenology. Section~\ref{sec:Analysis} presents the sensitivity analysis and resulting projected reach, and compares the results with existing experimental constraints. Finally, Section~\ref{sec:Conc} summarizes our results and conclusions.

\section{\label{sec:MAPP}The MoEDAL-MAPP Experimental Program}
\begin{figure}[!htb]
    \includegraphics[width = 8.5 cm]{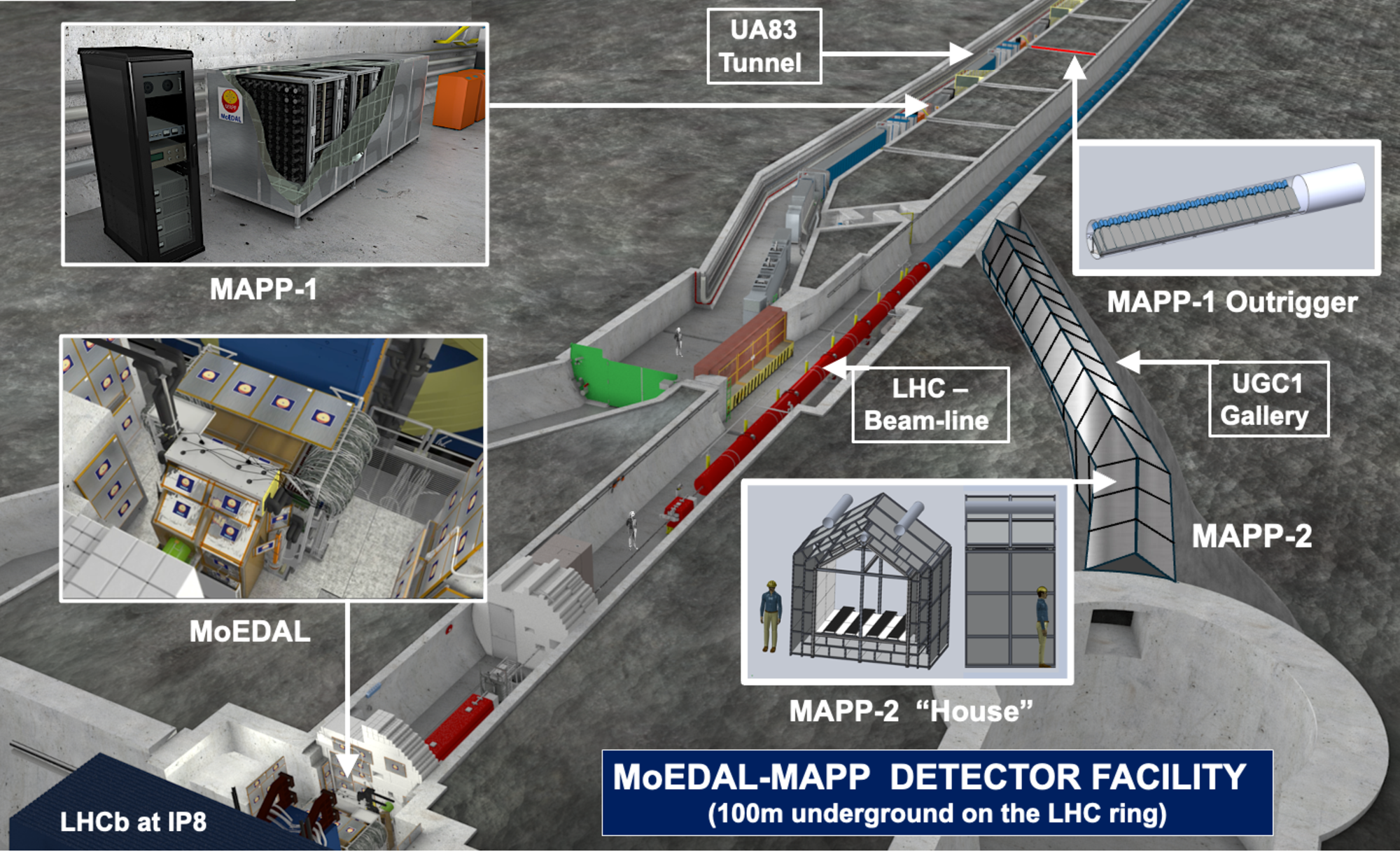}
    \caption{Schematic overview of the MoEDAL-MAPP facility at the LHC, indicating the locations of the MoEDAL detector at IP8, the Phase-1 MAPP Detector (MAPP-1) in the UA83 gallery, the MAPP-1 Outrigger Detector (MAPP-1 Outrigger) adjacent to MAPP-1, and the Phase-2 MAPP Detector (MAPP-2) in the UGC1 gallery. Reproduced with permission from Ref.~\cite{Staelens2024}.}
    \label{fig:MoEDAL-MAPP-Facility}
\end{figure}

The MoEDAL-MAPP experimental program comprises three detector systems: the Phase-1 MAPP Detector (MAPP-1), the MAPP-1 Outrigger Detector (MAPP-1 OD), and the Phase-2 MAPP Detector (MAPP-2). These detectors are optimized for complementary searches for FIPs over a broad range of masses and couplings, extending the discovery potential of the LHC into regions of parameter space inaccessible to conventional general-purpose detectors. The location of each MAPP detector system within the LHC infrastructure relative to Interaction Point 8 (IP8) is shown schematically in Fig.~\ref{fig:MoEDAL-MAPP-Facility}.

The following subsections provide descriptions of the three MoEDAL-MAPP detector systems. Brief overviews of MAPP-1 and the MAPP-1 OD are provided, while a detailed description of MAPP-2 is presented. Comprehensive descriptions of MAPP-1 and the MAPP-1 OD can be found in Refs.~\cite{Pinfold2021,Kalliokoski2024} and \cite{Pinfold2024,Kalliokoski2025}, respectively.

\subsection{\label{sec:MAPP1a}MAPP-1}
MAPP-1 is an active detector consisting of four collinear sections, each comprising $100$ scintillation bars individually coupled to photomultiplier tubes (PMTs), providing a sensitive volume of approximately $1.2$~m $\times$ $1.2$~m $\times$ $3$~m. It is located in the LHC's UA83 gallery downstream of IP8, approximately $100$~m from the interaction point at an angle of $7.3^{\circ}$ with respect to the beamline. The detector is hermetically enclosed within an active veto system to suppress backgrounds and further protected by a flame shield. Although its primary focus is the search for minicharged particles ($Q << e$)~\cite{Kalliokoski2024}, MAPP-1 can also operate as a displaced-vertex (DV) detector, searching for long-lived FIPs that decay within its sensitive volume into visible final states. Previous studies have demonstrated its sensitivity to unexplored regions of parameter space in a variety of such models~\cite{deVries2021,Dreiner2021,Popa2022,Bertuzzo2022,Beltran2023,Dreiner2023,Staelens2024,Alemany2025}. MAPP-1 was installed in the UA83 gallery during LHC Run~3; the detector is currently undergoing commissioning.

\subsection{\label{sec:MAPP1}The MAPP-1 Outrigger Detector}
The proposed MAPP-1 OD is an auxiliary detector designed to complement MAPP-1 by providing additional sensitivity to high-mass minicharged particles with electric charges as small as $\sim 0.01 \: e$~\cite{Kalliokoski2025}. Similar to MAPP-1, the MAPP-1 OD can also provide sensitivity to long-lived FIPs decaying into charged final states in the region upstream of or within the detector. It is planned for installation in Duct~4 of UA83, a former ventilation duct located between the MAPP-1 detector and the LHC beamline, approximately $120$~m downstream of IP8~\cite{Pinfold2024}. The detector comprises $80$ plastic scintillator slabs ($60$~cm $\times$ $30$~cm $\times$ $5$~cm) arranged in a four-layer staggered geometry, with each slab individually instrumented with a low-noise PMT. This layered hodoscope configuration provides coincidence-based event reconstruction, enabling background rejection and self-veto capabilities. The detector is assembled from ten modular units, each containing eight scintillator slabs mounted at a $45^{\circ}$ angle to increase the effective path length traversed by particles originating from the interaction point. Additionally, it is shielded from beam-related backgrounds by an iron absorber positioned at the beamline-facing end of the duct. The MAPP-1 OD readout and calibration systems are designed to share the same electronics architecture as MAPP-1. The Outrigger auxiliary detector will be considered for installation by the LHC Committee (LHCC) after the commissioning of the MAPP-1 detector has ended successfully.

\subsection{\label{sec:MAPP2}MAPP-2}
\begin{figure}[!htb]
    \includegraphics[width = 6.5 cm]{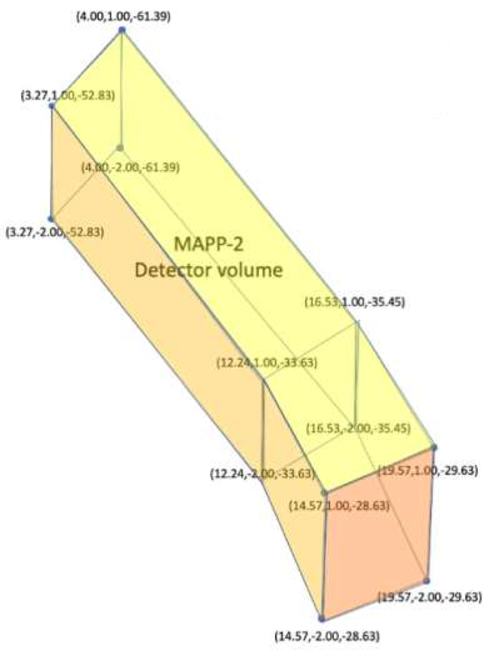}
    \caption{The proposed fiducial volume of the MAPP-2 detector. Dimensions indicated are in meters (maximum fiducial volume, $1400~\mathrm{m}^3$).}
    \label{fig:MAPP-2-Fiducial-Volume}
\end{figure}

The second MoEDAL Apparatus for Penetrating Particles (MAPP-2) is a proposed upgrade to the MoEDAL-MAPP program for deployment during the HL-LHC~\cite{pinfold2023moedalmapp}. It consists of a large instrumented decay volume in the UGC1 gallery adjacent to IP8, providing an approximately $1400~\mathrm{m}^3$ fiducial volume for searches for long-lived particles (LLPs). The proposed detector geometry and fiducial volume are shown in Fig.~\ref{fig:MAPP-2-Fiducial-Volume}. By instrumenting the gallery as a dedicated decay region, MAPP-2 is optimized to detect LLPs produced at IP8 and decaying into visible final states at macroscopic distances from the interaction point.

\begin{figure}[!htb]
    \includegraphics[width = 8.5 cm]{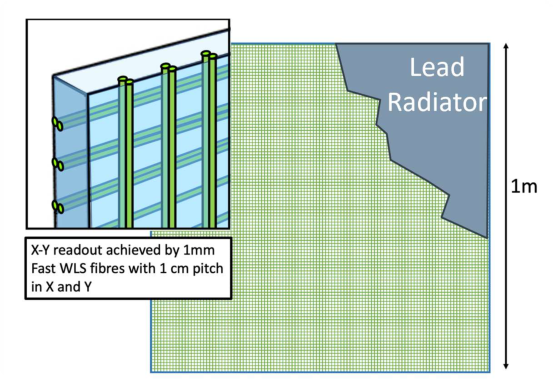}
    \caption{Schematic of a basic MAPP-2 scintillator tile; $X$--$Y$ position determination around $3$~mm in each coordinate. Reproduced with permission from Ref.~\cite{Staelens2024}.}
    \label{fig:MAPP-2-Tile-Schematic}
\end{figure}

The MAPP-2 detector is currently in the research and development phase. Its design comprises a layered geometry of large-area scintillator panels read out using fast wavelength-shifting (WLS) fibres coupled to silicon photomultipliers (SiPMs). The WLS fibres are arranged in orthogonal $X$--$Y$ configurations, enabling the reconstruction of charged-particle trajectories and displaced decay vertices. When fully instrumented, the fibre readout is expected to provide a position resolution of approximately $3~\mathrm{mm}$ in both spatial coordinates. A schematic representation of the MAPP-2 detector scintillator tile design is shown in Fig.~\ref{fig:MAPP-2-Tile-Schematic}.

The layered scintillator geometry enables the reconstruction of LLP decays through the identification of multiple displaced tracks and provides directional information for suppressing cosmic-ray backgrounds. Additionally, a lead radiator layer is incorporated to allow the detection of photons produced in LLP decays (e.g., from decays of photophilic axion-like particles), extending the range of accessible visible final states and broadening the potential LLP search program.

\begin{figure}[!htb]
    \includegraphics[width = 8.5 cm]{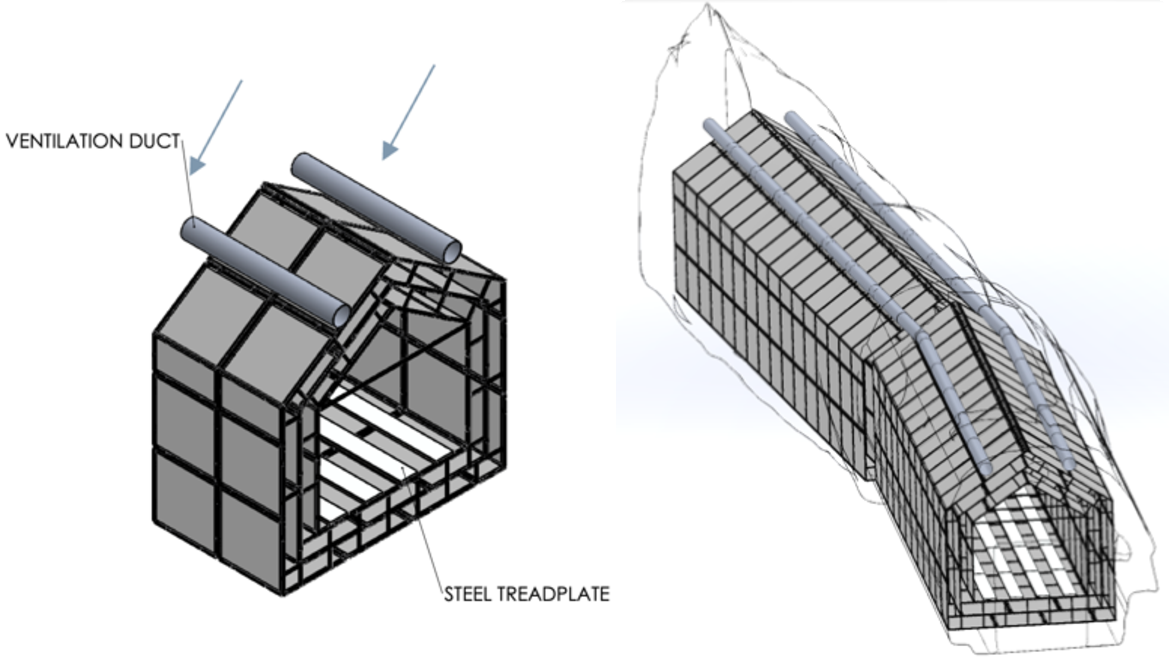}
    \caption{MAPP-2 detector construction subunit (left). Basic subunits combined to form MAPP-2, joined to fill the UGC1 gallery (right).}
    \label{fig:MAPP-2-schematics}
\end{figure}

A basic MAPP-2 detector subunit, shown in Fig.~\ref{fig:MAPP-2-schematics} (left), consists of three nested layers of scintillator tiles housed within a supporting steel structure. Two ventilation ducts are mounted along the top of the structure, while steel treadplates form the bottom surface. The full MAPP-2 detector, shown in Fig.~\ref{fig:MAPP-2-schematics} (right), is assembled from $14$ such subunits.

Deployment of MAPP-2 requires civil engineering work to prepare the UGC1 gallery for use as an experimental facility. Including this work, the total cost of MAPP-2 is estimated to be approximately USD~$3$--$4$~million. This relatively modest cost, combined with the large fiducial volume and complementary sensitivity, makes MAPP-2 an attractive option for LLP searches.

MAPP-2 is designed to provide highly cost-effective sensitivity complementary to existing and proposed searches for LLPs. A Notice of Intent for MAPP-2 has been presented to the LHCC, and a more detailed Expression of Interest is currently under preparation.

\section{\label{sec:HiggsPortal}Higgs-Portal Dark Sector and LHC Phenomenology}
We consider a minimal extension of the SM in which a dark scalar (dark Higgs) field, $S$, is introduced as part of a dark sector. The dark scalar field mixes with the SM Higgs field through a Higgs-portal quartic scalar interaction, as originally proposed in Ref.~\cite{Patt2006}. The corresponding scalar-sector Lagrangian can be written as
\begin{equation}
    \mathcal{L} = \mathcal{L}_{\mathrm{Kin}} + \mathcal{L}_{\mathrm{DS}} + \mu_{S}^{2} S^{2} - \frac{\lambda_{S}}{4} S^{4} + \mu^{2}|H|^{2} - \lambda |H|^{4} - \epsilon_{h} S^{2} |H|^{2}   \label{eq:lagrangian1}
\end{equation}
where $S$ is a real scalar field associated with the hidden sector, $H$ is the SM Higgs doublet, $\lambda_S$ and $\lambda$ are quartic scalar couplings, $\mu_S$ and $\mu$ are mass parameters, and $\epsilon_h$ is the Higgs-portal coupling. Here, $\mathcal{L}_{\mathrm{Kin}}$ and $\mathcal{L}_{\mathrm{DS}}$ denote the kinetic and any remaining dark-sector terms, respectively. The final term in Eq.~\eqref{eq:lagrangian1} describes the quartic interaction between the dark scalar and the SM Higgs field. The SM Higgs potential is given by
\begin{equation}
V \left( H^\dagger H \right) = -\mu^2 H^\dagger H + \lambda \left( H^\dagger H \right)^2.
\end{equation}

The scalar potential is invariant under the discrete symmetry $S \rightarrow -S$. After spontaneous symmetry breaking, both $S$ and $H$ acquire non-zero VEVs, denoted by $v_S$ and $v$, respectively. Minimizing the scalar potential with respect to the two fields yields
\begin{equation}
v_S^2 = \frac{2 \lambda \mu_S^2 - \epsilon_h \mu^2}
{\lambda \lambda_S - \epsilon_h^2},
\label{eq:vev_S}
\end{equation}
and
\begin{equation}
v^2 = \frac{\lambda_S \mu^2 - 2 \epsilon_h \mu_S^2}
{\lambda \lambda_S - \epsilon_h^2}.
\label{eq:vev_h}
\end{equation}

The resulting scalar mass matrix is obtained by expanding the fields about their respective VEVs,
\begin{equation}
H =
\frac{1}{\sqrt{2}}
\begin{pmatrix}
0\\
v + h_{\mathrm{SM}}
\end{pmatrix},
\qquad
S = v_S + s.
\end{equation}
The scalar mass matrix is diagonalized by a rotation with mixing angle $\theta$, yielding the physical mass eigenstates. The interaction and mass eigenstates are related through
\begin{equation}
h_{\mathrm{SM}} = \phi \sin\theta + h \cos\theta, \qquad
s = \phi \cos\theta - h \sin\theta,
\end{equation}
where $\phi$ is the $CP$-even dark scalar and $h$ is the $125$~GeV SM-like Higgs boson. The mixing angle is given by
\begin{equation}
\begin{aligned}
\tan\theta &=
\frac{2 \epsilon_h v v_S}
{\lambda v^2 - \lambda_S v_S^2
+ \sqrt{(\lambda v^2 - \lambda_S v_S^2)^2
+ 4 \epsilon_h^2 v^2 v_S^2}} \\
& \simeq \frac{2 \epsilon_h v v_S}{m_h^2 - m_\phi^2},
\end{aligned}
\end{equation}
where the latter approximation applies for small portal coupling $\epsilon_h$ and $m_{\phi} \neq m_h$.

In the mass basis, the resulting effective interactions are
\begin{equation}
\mathcal{L}_{\mathrm{eff}} =
- \frac{1}{2} m_{\phi}^{2} \phi^{2}
- \sin\theta \frac{m_f}{v} \phi \bar{f} f
- \epsilon_{h} v h\phi^{2},
\label{eq:lagrangian2}
\end{equation}
where $m_{\phi}$ and $m_f$ denote the dark scalar and fermion masses, respectively. The mixing therefore induces couplings between $\phi$ and the SM fermions proportional to the SM Yukawa couplings and $\sin\theta$. Due to the $\theta$ suppression, the dark scalar can exhibit long decay lengths across a substantial region of the viable $(\theta, m_\phi)$ parameter space. Higher-order terms in $h$ and $\phi$ have been neglected. The SM Higgs VEV, $v\simeq246$~GeV, and mass, $m_h\simeq125$~GeV, are fixed by experiment, leaving $m_\phi$ and $\theta$ as the relevant free parameters for the phenomenological analysis.

\begin{figure}[!!htb]
    \includegraphics[width = 8.5 cm]{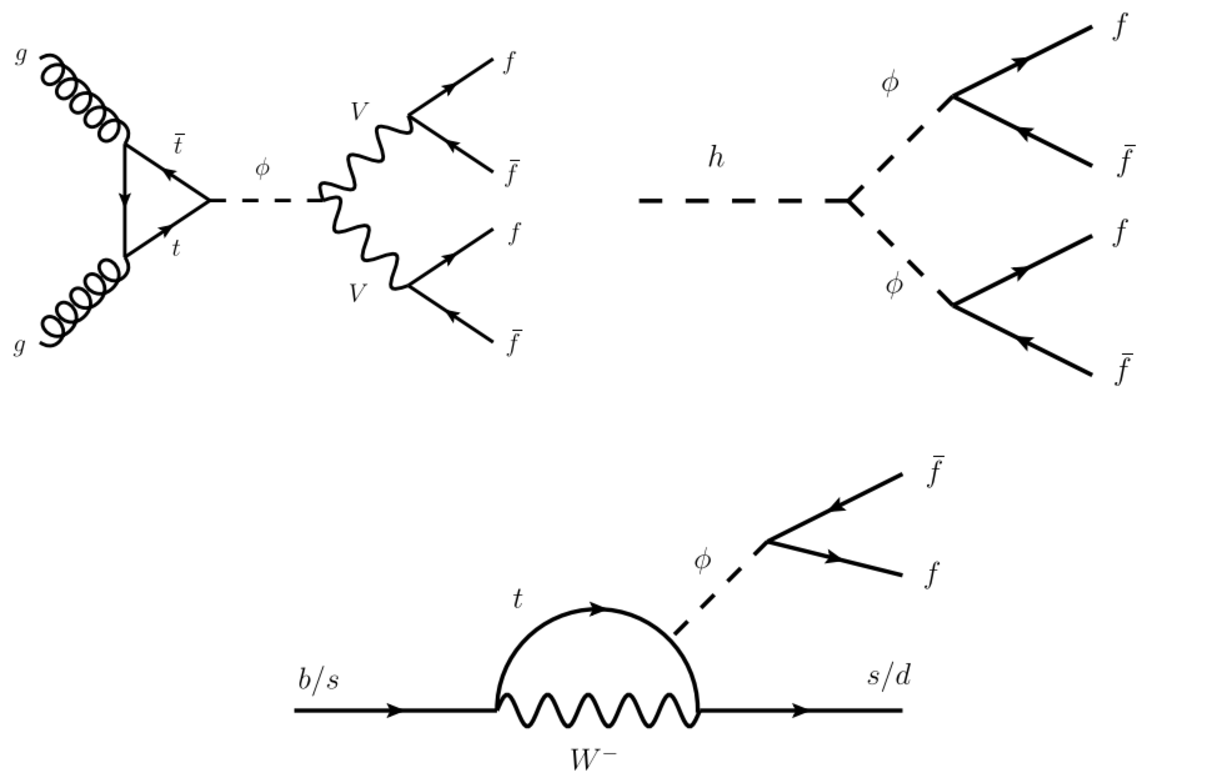}
    \caption{Feynman diagrams illustrating three representative dark scalar production modes at colliders with subsequent decays to fermionic final states: (upper left) gluon fusion, (upper right) SM Higgs decays, and (lower) rare $K$ and $B$ decays. Adapted from Ref.~\cite{Ferber2024}.}
    \label{fig:DH-FDs}
\end{figure}

We consider two phenomenologically relevant production mechanisms for the dark scalar at the LHC: inclusive $B \rightarrow X_s\phi$ decays and rare Higgs decays, $h \rightarrow \phi\phi$. Fig.~\ref{fig:DH-FDs} shows the corresponding Feynman diagrams including subsequent $\phi$ decays to fermionic final states. We restrict our study to muonic decays of the dark scalar, $\phi \rightarrow \mu^+\mu^-$, providing a clean visible final state.

The corresponding branching ratios are obtained as follows. For the Higgs decay, we derive the branching ratio for $h \rightarrow \phi \phi$ in the small-$\theta$ limit as 
\begin{equation}
    \mathcal{B} \left( h \rightarrow \phi \phi \right) \simeq \frac{v^2 \epsilon_h^2}{8 \pi m_h \Gamma_h} \sqrt{1 - \frac{4 m_{\phi}^2}{m_h^2}}
        \label{eq:BR-htophiphi}
\end{equation}
where $\Gamma_h$ is the total Higgs decay width. For Higgs-mediated dark scalar production, we adopt the branching fraction $\mathcal{B} \left( h\rightarrow \phi \phi \right) = 0.01$, following the phenomenological benchmarks in Ref.~\cite{Alemany2025}. This choice provides a convenient reference point for comparing the projected sensitivities of MAPP-1 and MAPP-2 with those of other experiments. The branching fraction depends on the parameters of the scalar potential and is therefore treated here as an independent phenomenological benchmark rather than a fixed theoretical prediction. The resulting sensitivity projections are conditional on this assumption, and the expected Higgs-mediated signal yield scales linearly with $\mathcal{B} \left( h\rightarrow \phi \phi \right)$ at fixed scalar mass and mixing angle. For the $B \rightarrow X_{s} \phi$ production mode, we use the following branching ratio~\cite{Willey1982, Chivukula1988, Grinstein1988, PDGrev2024, Gligorov2018}:
\begin{equation}
    \mathcal{B} \left( B \rightarrow X_{s} \phi \right) = 6.2 \left( 1 - \frac{m_\phi^2}{m_b^2} \right) \sin^2\theta.
    \label{eq:BR-BtoXsphi}
\end{equation}

\section{\label{sec:Analysis}Projected Sensitivity}
\begin{figure}[!htb]
    \includegraphics[width = 8.5 cm]{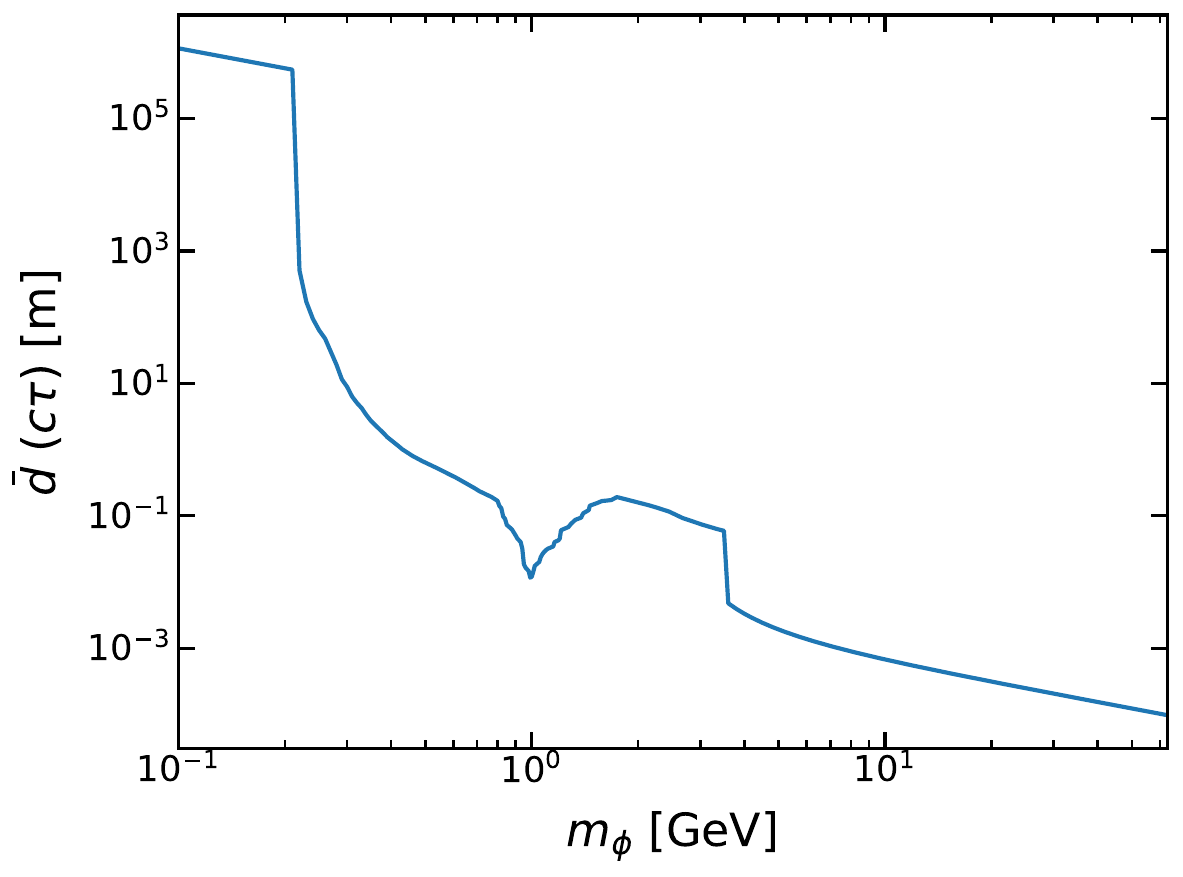}
    \caption{Dark scalar proper decay length as a function of mass ($\sin\theta = 0.01$).}
    \label{fig:dark-scalar-lifetime}
\end{figure}

We simulate dark scalar production in $pp$ collisions at $\sqrt{s}=14$~TeV through rare $B$ and Higgs-boson decays using \textsc{Pythia}~8 (v.~8.314)~\cite{Sjostrand2015, Bierlich2022}, with the default Monash tune~\cite{Skands2014}. Specifically, we consider $B \rightarrow K \phi$ as a proxy for the inclusive $B \rightarrow X_s\phi$ process to generate the dark scalar kinematics, while using the inclusive $B \rightarrow X_s\phi$ rate from Eq.~\ref{eq:BR-BtoXsphi} for normalization. We also consider Higgs decays $h \rightarrow \phi \phi$. Over the mass range of interest, the dark scalar lifetime is complicated, requiring different treatments across the relevant mass regimes. We therefore follow the approach of Ref.~\cite{Feng2018}. For light dark scalars with $m_{\phi} < 2 m_{\pi}$, the dominant decay modes are leptonic, primarily $\phi \rightarrow e^{+} e^{-}$ and $\phi \rightarrow \mu^{+} \mu^{-}$. The corresponding decay width is
\begin{equation}
\Gamma(\phi\rightarrow \ell^+\ell^-) =
\frac{m_\ell^2\theta^2}{v^2}\frac{m_{\phi}}{8\pi}
\left(1-\frac{4m_\ell^2}{m_{\phi}^2}\right)^{3/2},
\qquad \ell=e,\mu.
\end{equation}
Between $2m_\pi < m_\phi \lesssim 1$~GeV, we use the numerical results of Ref.~\cite{Bezrukov2013}, which are partially based on Ref.~\cite{Donoghue1990}. In the range $1$~GeV $\lesssim m_\phi \lesssim 2.5$~GeV, we calculate the decay width using~\cite{Winkler2019}
\begin{equation}
\Gamma =
\frac{1}{4 \pi \sqrt{2}} \sin^2\theta G_F m_u^2 m_\phi
\left(1-\frac{4 m_u^2}{m_\phi^2}\right).
\end{equation}
At higher masses, $m_\phi \gtrsim 2.5$~GeV, we use the spectator model~\cite{Gunion2000,McKeen2009}. The resulting proper decay length, $c \tau$, of the dark scalar is shown in Fig.~\ref{fig:dark-scalar-lifetime} for $\sin\theta = 0.01$. The characteristic dip near $m_\phi = 1$~GeV arises from mixing between the dark scalar and the $f_0 \left(980\right)$ resonance, which affects the adopted hadronic-width treatment and consequently the decay acceptance.

We use the dark scalar lifetime calculated following this approach as an input to \textsc{Pythia} to simulate its decay, considering the $\phi \rightarrow \mu^+\mu^-$ channel and assuming a $100\%$ branching ratio to this final state. For this sensitivity study, we use the decay $\phi \rightarrow \mu^+\mu^-$ as a representative visible two-track final state for assessing the detectability of long-lived dark scalars in MAPP-1 and MAPP-2. The branching fractions of dark scalars into individual final states depend on the underlying model and scalar mass. To provide a simple benchmark normalization of the visible-decay sensitivity and facilitate comparison with other phenomenological studies, we therefore adopt the assumption $\mathcal{B} \left( \phi \rightarrow \mu^+\mu^- \right) = 1$. This assumption is not intended to represent the physical dimuon branching fraction of a particular Higgs-portal model, but rather an idealized benchmark for visible two-track decays. The resulting projections should consequently be interpreted as optimistic sensitivity estimates, with the sensitivity to specific decay modes depending on their physical branching fractions and reconstruction efficiencies; sensitivities for smaller branching fractions would be correspondingly reduced.

\begin{figure}[!htb]
    \includegraphics[width = 8.5 cm]{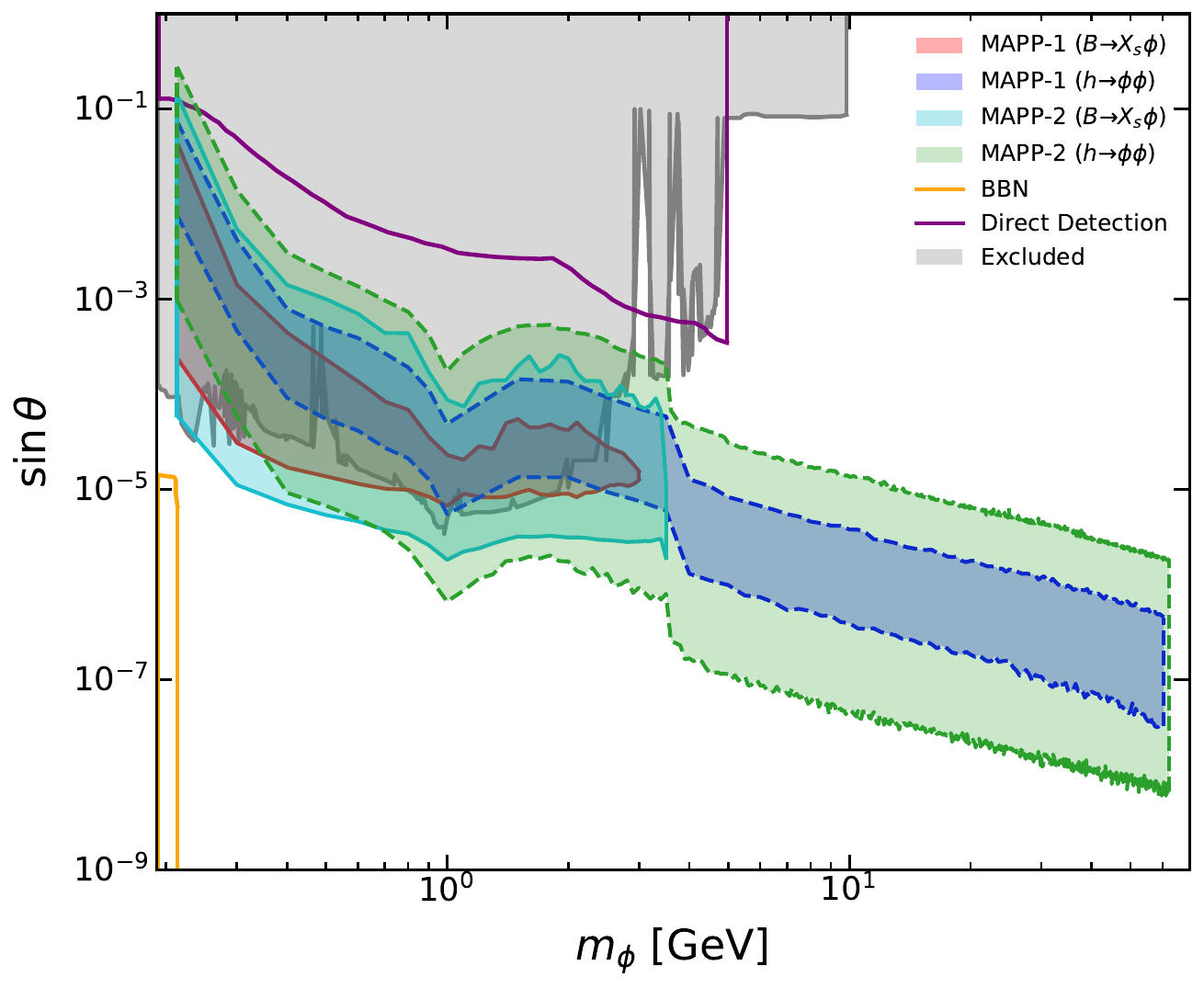}
    \caption{Projected $95\%$ confidence level (CL) sensitivities of MAPP-1 and MAPP-2 for dark scalars produced in $14$~TeV $pp$ collisions, with the contributions from rare $B$ and $h$ decays shown separately. The projections assume a background-free search with $100\%$ signal efficiency, a total integrated luminosity of $300$~fb$^{-1}$, and $\mathcal{B} \left( h\rightarrow \phi \phi \right) = 0.01$. The excluded region comprises the following existing constraints: at $90\%$~CL, PS191~\cite{GORBUNOV2021}, E949~\cite{Artamonov2009}, NA62~\cite{CortinaGil2021-1, CortinaGil2021-2}, KOTO~\cite{Ferber2024, Ahn2021}, KTeV~\cite{Alavi-Harati2000}, and BaBar~\cite{Winkler2019, Lees2015}; at $95\%$~CL, MicroBooNE~\cite{Ferber2024, Abratenko2021, Abratenko2022}, L3~\cite{Ferber2024, ACCIARRI1996}, CHARM~\cite{Winkler2019, BERGSMA1985}, LHCb~\cite{Winkler2019, Aaij2015, Aaij2017}, Belle~II~\cite{Adachi2023, Adachi2025}, and CMS~\cite{Tumasyan2022}. Constraints from Big Bang nucleosynthesis (BBN) for $\tau_{\phi} > 0.1~\mathrm{s}$~\cite{Ferber2024} and direct detection assuming $m_{\rm DM} = 2m_\phi$~\cite{Ferber2024} are shown separately.} 
    \label{fig:MAPP-proj-sens-sep-ch}
\end{figure}

\begin{figure}[!htb]
    \includegraphics[width = 8.5 cm]{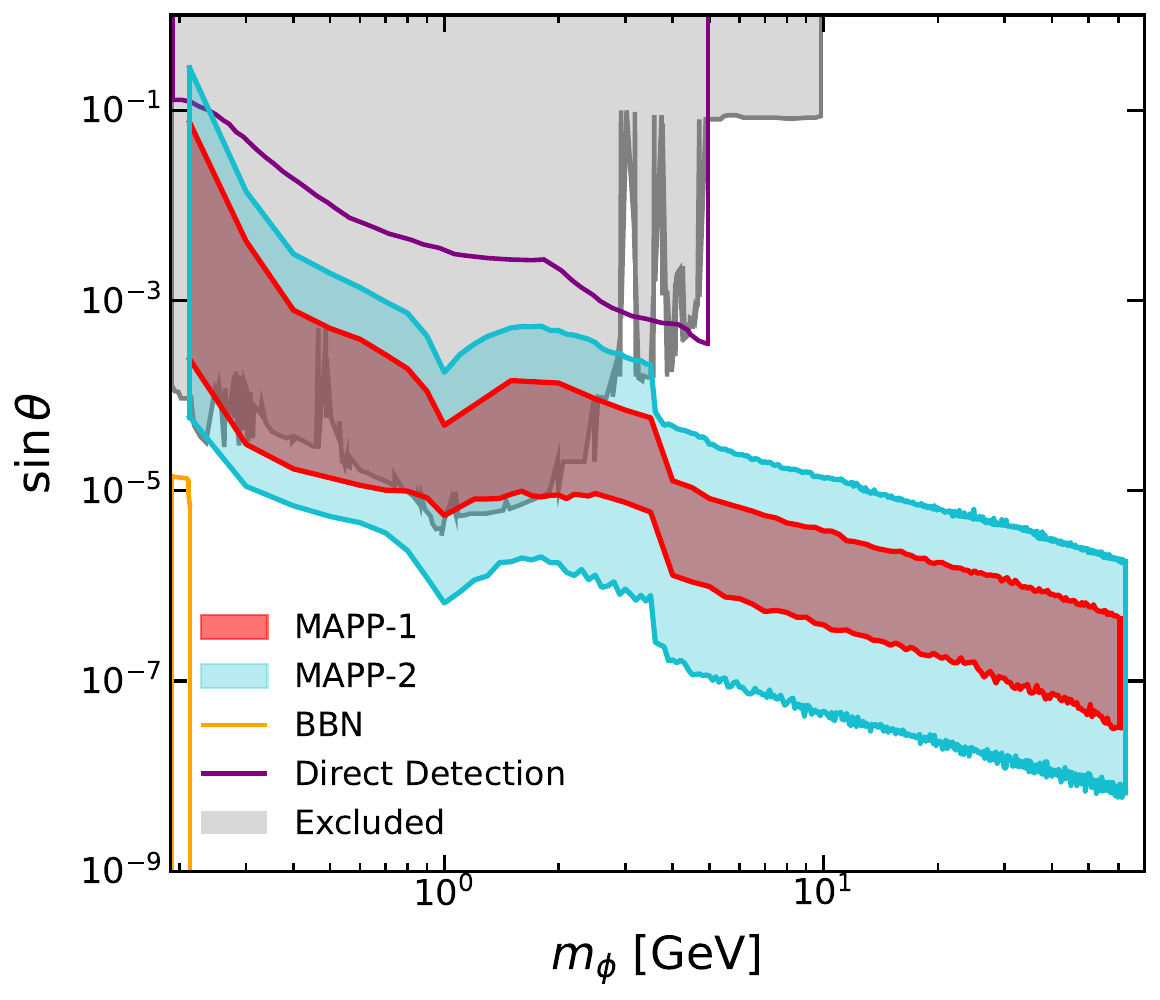}
    \caption{Projected $95\%$ confidence level (CL) sensitivities of MAPP-1 and MAPP-2 for dark scalars produced in $14$~TeV $pp$ collisions via rare $B$ and $h$ decays, assuming a background-free search with $100\%$ signal efficiency, a total integrated luminosity of $300$~fb$^{-1}$, and $\mathcal{B} \left( h\rightarrow \phi \phi \right) = 0.01$. The excluded region comprises the following existing constraints: at $90\%$~CL, PS191~\cite{GORBUNOV2021}, E949~\cite{Artamonov2009}, NA62~\cite{CortinaGil2021-1, CortinaGil2021-2}, KOTO~\cite{Ferber2024, Ahn2021}, KTeV~\cite{Alavi-Harati2000}, and BaBar~\cite{Winkler2019, Lees2015}; at $95\%$~CL, MicroBooNE~\cite{Ferber2024, Abratenko2021, Abratenko2022}, L3~\cite{Ferber2024, ACCIARRI1996}, CHARM~\cite{Winkler2019, BERGSMA1985}, LHCb~\cite{Winkler2019, Aaij2015, Aaij2017}, Belle~II~\cite{Adachi2023, Adachi2025}, and CMS~\cite{Tumasyan2022}. Constraints from Big Bang nucleosynthesis (BBN) for $\tau_{\phi} > 0.1~\mathrm{s}$~\cite{Ferber2024} and direct detection assuming $m_{\rm DM} = 2m_\phi$~\cite{Ferber2024} are shown separately.} 
    \label{fig:MAPP-proj-sens-comb}
\end{figure}

Based on the dark scalar lifetime and kinematics for each production mode, we determine the decay position and corresponding fiducial acceptance of MAPP-1 and MAPP-2. The simulations are performed over the parameter space of interest, $2m_\mu \lesssim m_\phi \lesssim m_h/2$, with $\sin\theta$ varied for each sampled value of $m_\phi$. For each point in the parameter space and for each production mode, we independently determine the number of $\phi$ decays occurring within the fiducial volumes of MAPP-1 and MAPP-2, with the expected signal yield given by $N_{\mathrm{sig}} = \mathcal{L}_{\mathrm{int}} \, \sigma_{\mathrm{prod}} \, \mathcal{B}_{\mathrm{prod}\to\phi+X} \, \mathcal{B}(\phi \rightarrow \mu^+\mu^-) \, \epsilon_{\mathrm{sig}} \, A_{\mathrm{fid}}$. Here, $N_{\mathrm{sig}}$ is the estimated number of signal events, $\sigma_{\mathrm{prod}}$ is the production cross-section for either $B$ mesons or Higgs bosons, $\mathcal{B}_{\mathrm{prod}\to\phi+X}$ is the corresponding branching ratio for dark scalar production (Eq.~\ref{eq:BR-htophiphi} or Eq.~\ref{eq:BR-BtoXsphi}), $\mathcal{B}(\phi \rightarrow \mu^+\mu^-)$ is the dark scalar dimuon branching ratio, $\epsilon_{\mathrm{sig}}$ is the overall detector efficiency (assumed to be $100\%$), and $A_{\mathrm{fid}}$ is the detector fiducial acceptance. For Higgs boson production via gluon fusion, we use a production cross-section of $\sigma(gg\rightarrow h)=54.67$~pb, while for $b\bar{b}$ production, we assume a cross-section of $500$~$\upmu$b. Projected sensitivity limits are then obtained using the $CL_{s}$ method~\cite{Read2002}, assuming a total integrated luminosity of $300$~fb$^{-1}$, a background-free scenario, and $100\%$ signal efficiency. Under these assumptions, the expected background-free $95\%$ CL sensitivity corresponds to approximately $3.0$ expected signal events. The projected sensitivities for the $B$-meson decay and Higgs-mediated production modes are shown separately for MAPP-1 and MAPP-2 in Fig.~\ref{fig:MAPP-proj-sens-sep-ch}, alongside the current exclusion bounds, Big Bang nucleosynthesis (BBN) constraints, and direct-detection limits. The sensitivities from the two production modes are then combined to obtain a single sensitivity curve for each detector, as shown in Fig.~\ref{fig:MAPP-proj-sens-comb}, together with the same constraints.


The projected sensitivities of MAPP-1 and MAPP-2 extend into significant regions of parameter space that remain largely unexplored or unconstrained by existing searches. In particular, both detectors probe small mixing angles corresponding to macroscopic dark scalar decay lengths, where the sensitivity of conventional general-purpose detector searches is reduced or absent. MoEDAL-MAPP therefore provides a complementary probe of dark scalars.

A key limitation of this study is the assumption of negligible backgrounds and $100\%$ signal efficiency. Incorporating realistic backgrounds (e.g., decays of $K_L^0$ and secondary hadrons) and detector response is therefore an important direction for future work and is currently ongoing. These effects will be studied using the detailed \textsc{Geant4} model of the MoEDAL-MAPP facility, known as SUMMA (Simulation of the UA83, MoEDAL, MAPP Arena)~\cite{Kalliokoski2024c}. Future studies could also consider additional dark scalar production modes, which may further enhance the potential sensitivity of the MoEDAL-MAPP experiment.


\section{\label{sec:Conc}Conclusions}
The MoEDAL-MAPP program provides a dedicated experimental facility for feebly interacting particle searches at the LHC. In this work, we investigated the sensitivity of the MAPP-1 and MAPP-2 detectors at the HL-LHC to Higgs-portal dark scalars as candidate LLPs. We simulated dark scalar production through rare $B$ decays and exotic Higgs decays and determined the resulting sensitivity of both detectors.

Our results demonstrate that the MoEDAL-MAPP program provides a complementary probe of Higgs-portal dark scalars at the HL-LHC. Under background-free and $100\%$ signal efficiency assumptions, and assuming $\mathcal{B} \left( h\rightarrow \phi \phi \right) = 0.01$, MAPP-1 probes dark scalar masses from $2m_\mu \lesssim m_\phi \lesssim m_h/2$ and mixing angles down to $\sin\theta\sim10^{-7}$, while MAPP-2 reaches mixing angles as small as $\sin\theta\sim10^{-8}$ over the same mass range. These results provide the first published LLP sensitivity projections for MAPP-2 incorporating Higgs-mediated dark-scalar production and demonstrate its potential to probe long-lived dark scalars with lifetimes beyond the reach of conventional general-purpose collider searches. 

More broadly, Higgs-portal scalars provide a minimal, well-motivated mediator between the Standard Model and a hidden sector, and arise naturally in scenarios addressing both dark matter and cosmic inflation. By extending sensitivity to such feebly-coupled states over unprecedented lifetimes, MAPP-1 and MAPP-2 position the MoEDAL-MAPP program to make a valuable contribution to the broader search for new physics beyond the Standard Model.

\begin{acknowledgments}
We acknowledge support by NSERC via Grant Number SAPPJ-2023-00043. M.S. acknowledges support by the Generalitat Valenciana via the APOSTD Grant No.\ CIAPOS/2021/88 and internal research funding provided by the University of Regina. We also thank the authors and maintainers of \textsc{Pythia}~8.
\end{acknowledgments}



\bibliography{DarkScalars}

\end{document}